\documentclass{cernrep} 
\usepackage{texnames}
\usepackage[T1]{fontenc}
\begin{document}
\title{Support Vector Machines in High Energy Physics}
 
\author{Anselm Vossen}

\institute{Albert-Ludwigs-Universit\"at, Freiburg, Germany}

\maketitle 

\begin{abstract}
This lecture will introduce the Support Vector algorithms for classification and regression.
They are an application of the so called kernel trick, which allows the extension 
of a certain class of linear algorithms to the non linear case.
The kernel trick will be introduced and in the context of structural risk minimization, large margin 
algorithms for classification and regression will be presented.
Current applications in high energy physics will be discussed.
\end{abstract}

\section{Introduction}
Multivariate analysis has become an essential tool in high energy physics analysis.
Widely used algorithms include Neural Networks (NNs) and boosted decision trees (BDTs). A fairly new
development are Support Vector Machines (SVMs) for classification and, closely related, Support Vector Regression (SVR).
These algorithms are competitive with NNs and BDTs in their predictive power but have the advantage 
that they are firmly rooted in statistical learning theory. 
This lecture will mainly address the more popular SVMs and discuss only briefly SVR, 
since classification problems are more common and the basic principles are similar. 
SVMs combine the linear maximum margin classifier, which is motivated by structural risk minimization with
the so-called kernel trick. 
The kernel trick allows the construction of a non-linear classifier by using a non-linear measure of
similarity between feature vectors which can be adapted to the problem at hand. 
It is a powerful concept that is not limited to the use in SVMs but can be used for any linear algorithm
that fulfills certain constraints. The advantage being that, once a kernel is developed, it can be used for 
any kernelized algorithm.
The design of kernels for specific problems is a field of research that is very much active, so one can expect
further improvements. 
This is another reason to have a look at the theory and the application in HEP.
The growing popularity has led to the availability of easy-to-use tools, which will be described in section \ref{toolSect}. 
Among others, an implementation in the ROOT toolkit for multivariate analysis TMVA \cite{tmva} is now available.
This lecture will first introduce the basic concepts of support vector machines in the linear case and the extension
to the nonlinear case with the kernel trick.
Then the connection to structural risk minimization and present applications and tools in high energy physics will be presented.

\section{Linear Classification}
\begin{figure}
\centerline{\includegraphics[width=5cm]{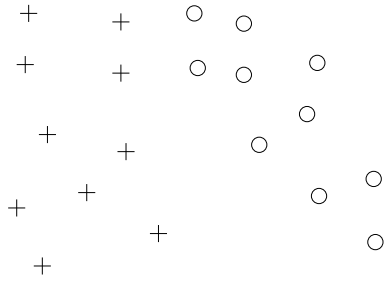}}
\caption{Two class problem: separate crosses and circles}
\label{twoClassFig}
\end{figure}

\begin{figure}
\centerline{\includegraphics[width=5cm]{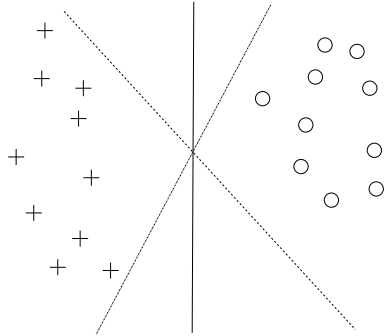}}
\caption{Several possibilities to solve the two class problem}
\label{figManyLines}
\end{figure}

This section introduces linear classification to solve a two-class classification problem, illustrated in figure \ref{twoClassFig}.
In this case it is possible to separate the two classes, dots and crosses, by a line. 
It is thus called a separable problem. Not all problems are separable, that is 
why the extensions to non-separable problems will be discussed later on.
As shown in the chapter about basic classification, this is a problem of supervised learning. 
A set of feature vectors (fv) $\vec{x} \in H$ is given, where $H$ denotes the feature space that contains the fvs.
The goal is now to construct a function $f: H\rightarrow {\pm 1 }$, that assigns a class label to a  given feature vector.
Fig. \ref{figManyLines} depicts a two dimensional instance of the problem.
There are many different ways of constructing this function, even with the restriction to linear functions.
The task is to find a function - in the two dimensional case a line, in more dimensions a hyperplane - that 
is optimal in a way. 

\begin{figure}
\centerline{\includegraphics[width=7cm, trim=50 200 0 0, clip]{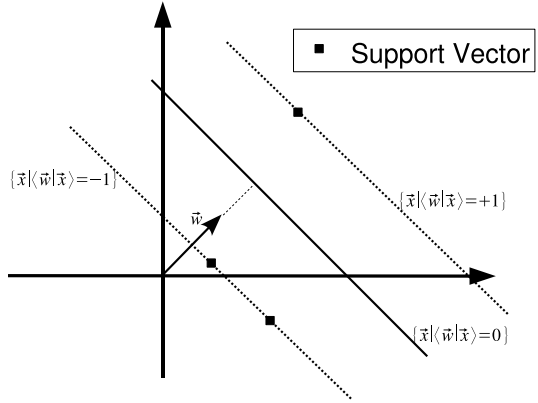}}
\caption{Maximum margin and support vectors}
\label{figWithSVs}
\end{figure}

The optimality criterion that will be used in the following is a maximal margin to the closest points of either class.
This is the so called maximum margin classifier. 
The reasoning behind this optimality criterion is that a wide margin corresponds to a good generalizing ability of the resulting classifier.
In order for a linear classifier to work at all, the instances of each class have to be concentrated on one side of the hyperplane. 
Having the border as far away as possible from the training examples intuitively maximizes the probability that instances of each class that have
not yet been observed, are classified correctly because they are expected to be in the neighborhood of the training examples of the same class.
Also, mathematically, one can show that the size of the margin is directly related to the generalizing power of 
a function. This feature is presented later in the context of structural risk minimization and Vapnik-Chervonenkis
theory.
After the definition of the optimality criterion, a way has to be found to construct this separating hyperplane.
One observation is that the hyperplane is completely determined by the feature vectors of both classes that
are the closest. They are called Support Vectors. This is because they ``support'' the separating hyperplane.
In an image motivated by physics, they are exerting a force that pushes from both sides on the plane and thus leads to a maximum margin to both classes.
The hyperplane can be described by the equation $\left<\vec{w},\vec{x}\right>+b=0$ where all 
vectors $\vec{x}$ that fulfill the equation lie on the plane and $\vec{w}$ and $b$ have to be 
determined.
Since the scale is arbitrary, and, for the sake of convention, the examples $\vec{x}_1$ of the one
class are classified by
 \begin{equation}
\label{margin1}
\left<\vec{w},\vec{x}_1\right> \le -1
\end{equation}
 and examples of the other class $\vec{x}_2$ by 
\begin{equation}
\label{margin2}
 \left<\vec{w},\vec{x}_2\right> \ge1,
\end{equation}
  two more conditions for 
examples that lie on the margin are constructed. Namely Eqs. (\ref{margin1}) and (\ref{margin2}).
This is illustrated in Fig. \ref{figWithSVs}. Here it can also be seen that from Eqs. 
(\ref{margin1}) and (\ref{margin2}) an expression for the size of the margin follows:
\begin{eqnarray}
\left<\vec{w},(\vec{x}^1-\vec{x}^2)\right>&=&2\\
\label{marginFormula}
\Rightarrow \left<\frac{\vec{w}}{\|\vec{w}\|},(\vec{x}^1-\vec{x}^2)\right>&=&\frac{2}{\|\vec{w}\|}
\end{eqnarray}
Finding the classification function
\begin{equation}
f(\vec{x})=\left<\vec{x},\vec{w}\right>+b
\end{equation}
is thus an optimization problem. Maximizing the margin given in eq. (\ref{marginFormula})
leads to the target function that needs to be minimized:
\begin{equation}
\label{eqTau}
\tau(\vec{w})=\frac{1}{2}\|\vec{w}\|^2
\end{equation}

Constrained optimization problems can be solved by the method of Lagrangian multipliers \cite{varCalc} where the function to be optimized and the 
$k$ constraints are incorporated into a new function, the Lagrangian, which is unconstrained but includes an unknown scalar variable for each constraint.
The stationary points of the constrained problem in $n$ can then be computed from the stationary points of the Lagrangian in $n+k$ variables.
Important sufficient optimality criteria are the Kuhn-Tucker saddle point conditions.
For details see \cite{nonLinProg}.
The Lagrangian that incorporates the target function $\tau$ and the constraints (\ref{margin1}), (\ref{margin2}) has the form
\begin{equation}
\label{firstLagrangian}
L(\vec{w},b,\vec{\alpha})=\frac{1}{2}-\sum_{i=1}^m \alpha_i\left(y_i(<\vec{x}_i,\vec{w}>+b)-1\right).
\end{equation}
It has to be minimized with respect to the primary variables $\vec{w}$ and $b$ and maximized with respect to the Lagrange multipliers $\alpha_i$.
A big advantage of the support vector algorithm is, that the optimization problem to be solved is convex.
This means that an optimal solution is guaranteed to be found,
in contrast to other algorithms such as neural networks that try to solve the equation iteratively until it stabilizes,
and there is no guarantee that it will stabilize nor that the found solution is optimal.

The solution can be written in terms of the training examples $\vec{x}_i$ and the Lagrange multipliers $\alpha_i$ as
\begin{equation}
\label{classFkt}
f(\vec{x})=sgn\left(\sum_{i=1}^m y_i \alpha_i <\vec{x},\vec{x}_i>+b\right)
\end{equation}
where the sign of the function indicates on which side of the margin the feature vector $\vec{x}$ lies.
Crucial for the working of the algorithm is that for examples where the conditions in Eqs. (\ref{margin1}),(\ref{margin2}) are overfulfilled, that is 
$y_i(<\vec{w},\vec{x}_i>+b)-1 \geq 0$ the corresponding $\alpha_i$ is zero, which can be shown by the Karush Kuhn Tucker (KKT) condition \cite{svmCite}.
Consequently only Lagrange multipliers $\alpha_i$ that correspond to examples on the margin contribute to (\ref{classFkt}).
These examples are the previously mentioned support vectors and one thus talks about the support vector expansion.
As expected only feature vectors which ``support'' the margin are contributing to its position. 
Without this property, the algorithm would be quite useless, since the computation time for the classification alone
 grows linearly in the number of feature vectors for which $\alpha_i$ is non-zero.

\subsection{Kernel Trick}
\label{kernelTrickSect}
\begin{figure}
\centerline{\includegraphics[width=10cm]{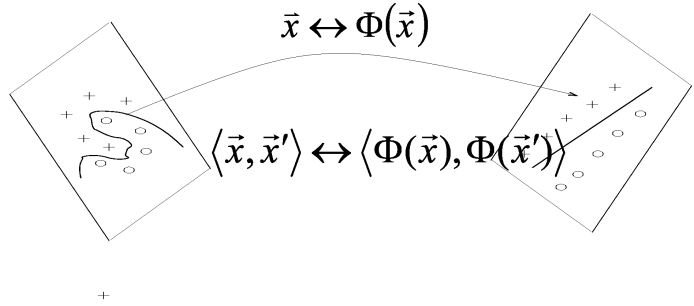}}
\caption{The kernel trick. Substituting the scalar product with the kernel function implies a scalar product of the mapped feature vectors}
\label{mappingFig}
\end{figure}

In the previous section a linear classifier has been constructed that has optimized generalizing abilities.
This section will introduce a method to extend this classifier to the nonlinear case.
This is done by mapping the feature vectors into another space $H'$, where they can be separated linearly by the existing algorithm.
In the original space the combination of mapping and linear separation leads to a non-linear classifier.
The space $H'$ has to be a Hilbert space of arbitrary dimensionality. In fact, the dimensionality can even be infinite, 
which would always allow a separation of the training examples.
The mapping function is denoted $\Phi$, the input space $H$ (for Hilbert) and the target space $H'$, thus
$\Phi: H \rightarrow H'$  (see illustration in Fig \ref{mappingFig}).

The mapping is inserted into the support vector expansion (Eq. \ref{classFkt}) which leads to an expression where the mapped feature and support vectors
only occur in scalar products, i.e. in the form $\left<\Phi(\vec{x}_i),\Phi(\vec{x})\right>$.
It follows that it is unnecessary to actually compute $\Phi$ for each vector, maybe augmenting the dimensionality, and then 
perform the computations. Instead it is only necessary to know the scalar product, which is a function $k:H\times H \rightarrow \mathbb R$.
It has been shown that it is actually enough to construct such a function $k$ that is positive definite (pd), 
so it is sufficient to show that the matrix $K_{ij}=k(\vec{x}_i,\vec{x}_j),\quad \vec{x}_i, \vec{x}_j \in H$ is pd for all $\vec{x}_i,\vec{x}_j$.
The existence of a $\Phi$ such that $k(\vec{x},\vec{x}')=<\Phi(\vec{x}),\Phi(\vec{x})`>$ is then guaranteed and the whole machinery that
has been developed for the linear case can be used. 
For interpretation purposes it is also possible to construct $\Phi$ explicitly from $k$ and the training examples \cite{svmCite}.
This is the content of Mercers Theorem \cite{mercerCite}, \cite{svmCite}.
Following this concept one can extend any linear algorithm that can be formulated in a way only depending on scalar products of the input vectors,
by replacing the scalar product with $k$.
This is known as the kernel trick and methods that make use of it as kernel methods.
Instead of choosing a mapping that is ideal for separation, with  the kernel trick an appropriate kernel function is chosen which is
much more intuitive, because the kernel function is like a measure for similarity between two feature vectors with respect to 
the class label. This makes it possible to choose this function according to the data and in this way adapt the algorithm to the given task.
In practice there are some well known kernel functions that perform well and that can also be used as a starting point 
to construct other kernels.
In the linear case the kernel function (or just kernel for short) is just the scalar product, other popular functions include 
\begin{itemize}
\item Gaussian kernel: $k(\vec{x},\vec{x}')=e^{-\gamma|\vec{x}-\vec{x}'|^2}$

\item Polynomial kernel: $k(\vec{x},\vec{x}')=\left(<\vec{x},\vec{x}'>+a\right)^d$

\item Linear kernel: $k(\vec{x},\vec{x}')=-|\vec{x}-\vec{x}'|$
\end{itemize}

\begin{figure}
\centerline{\includegraphics[width=3cm]{KFkt.eps}\hspace{3cm}\includegraphics[width=3cm]{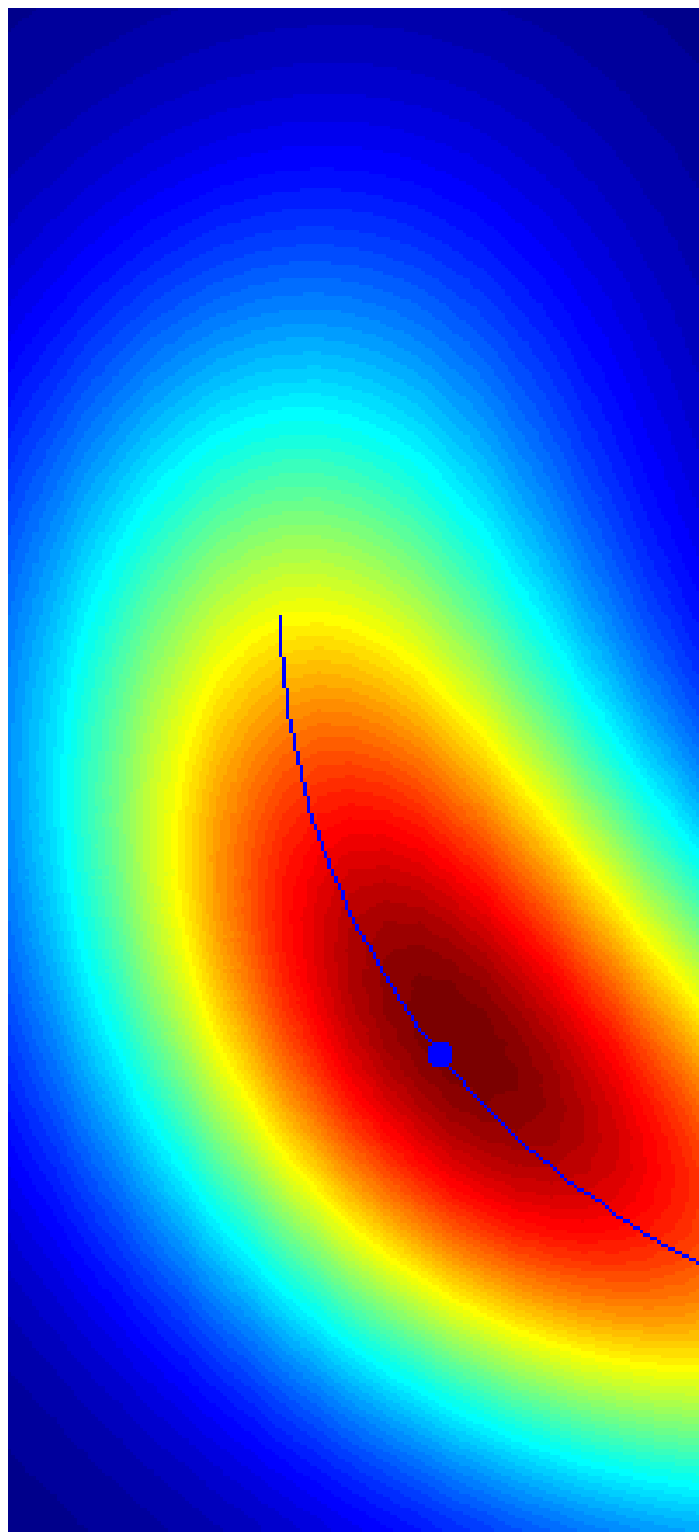}}
\caption{Illustration of the partial Haar-integration-kernel. Shown is $\vec{x}-\vec{x}'$ in the two dimensional case, unscaled. The dot indicates
\label{haarFig}
$\vec{x}=\vec{x}'$}
\end{figure}

Especially the Gaussian kernel is usually a good choice. It has also the advantage that there is only one parameter, $\gamma$,
which needs to be found. This can be done by scanning the parameter space and it is usually easy to find optimal parameters 
using $\gamma=0.1$ as a starting point and feature vectors which are normalized to the range $[-1,+1]$.
There exists also theoretical work dealing with the layout of the parameter space and subsequently the optimal parameter choice. 
In practice it is hardly a problem to find good parameters for the gaussian kernel.
This point will be revisited later in this lecture when the additional parameter $C$, that controls the generalizing ability has also been introduced.
An example for a more involved kernel is the Haar-integration-kernel, which uses integrations over a group, so called Haar-integration to 
construct a kernel that is invariant against transformations of the feature vector by group elements \cite{haarCite}.
Let $\mathcal{G}$ be a group of transformations on feature vectors $\vec{x} \in H$, $g, g'\in \mathcal{G}$ and $dg,dg'$ the measure, than  the Haar-integration-kernel $k_H$
is defined by
\begin{equation}
k_H(\vec{x},\vec{x}')=\int_\mathcal{G}\int_\mathcal{G} k_0(g\vec{x},g'\vec{x}')dgdg',
\end{equation}
where $k_0$ is a arbitrary kernel. The concept works also for partial integrations over a group, resulting in robustness against small 
transformations.
Figure \ref{haarFig} shows an example. There, a Gaussian kernel as the base kernel $k_0$ and the 
corresponding kernel making use of a partial Haar-integration are shown. 
The integration is performed over a partial circle, and equal colors indicate 
areas where the kernel function gives the same value. 
Which means areas in which the feature vectors have the same similarity.
In this two dimensional case the lines of equal similarity are concentric circles around $|\vec{x}-\vec{x}'|=0$, as 
expected from a Gaussian.
Once the integration is performed over a partial rotation it can be seen that the kernel function becomes
robust against small rotations of the feature vectors along the path of integration. 
This behavior can be used, for example, for the classification of images of numbers. There small rotations leave the
class unaltered, whereas a larger rotation can change a '6' to a '9'. 
With such a kernel it is thus possible to incorporate prior knowledge.
Kernel methods are a powerful concept, and once  a kernel is designed that is especially suited for a specific problem, 
it can be incorporated also in other kernel methods to extend linear methods to non linear ones. Examples are novelty detection or 
principal component analysis. More about specific kernels and their applications can be found in \cite{kernelMethodsCite}.

\section{Structural Risk Minimization - Soft Margin SVM}
This section will introduce the soft margin SVM. This formulation of the SVM algorithm with a so called soft margin
allows the application to the classification of non separable problems. In this case the conditions that all examples are classified correctly
cannot be incorporated in the  Lagrangian (\ref{firstLagrangian}).
Rather it has to be allowed, that an example lies on the wrong side
of the margin, but punished by a certain cost, that depends on the distance of the example from the margin. In the picture
of the support vectors pushing on the margin, this margin changes now from being hard, to a soft one, because the support vectors
are now allowed to push into it, and the force they are exerting on the margin becomes stronger when   they penetrate it deeper.
Allowing a soft margin will have an impact on the generalizing ability. The softer the margin the bigger the generalizing power of
the constructed classifier. This will give a link to structural risk minimization, which will be discussed in the following section. 
There the goal is to minimize the misclassification risk by finding a good trade-off between correctly classifying the training instances and
constructing a classifier that has a good generalizing ability.

\subsection{Structural Risk Minimization}
When designing a classifier the goal is always to minimize the expected risk, that is the expected misclassifications.
Given a probability distribution function (pdf) of feature vectors and classes $p(\vec{x},y)$ the expected risk is 
given by the expectation value of the error of the classification function $f$:
\begin{equation}
  R_{\mathrm{exp}}[f]=\int\left|f(\vec{x})-y\right|dp(\vec{x},y)
\end{equation}
 But what is available during training as the only quantity that can be used to estimate the expected risk with, 
 is the empirical risk. This is the error of the classification function during training given by:
\begin{equation}
R_{\mathrm{emp}}[f]=\frac{1}{m}\sum_{i=1}^m \frac{1}{2}\left|f(\vec{x})_i-y_i\right|
\end{equation}

\begin{figure}
\centerline{\includegraphics[width=5.5cm, trim=100 180 170 250, clip]{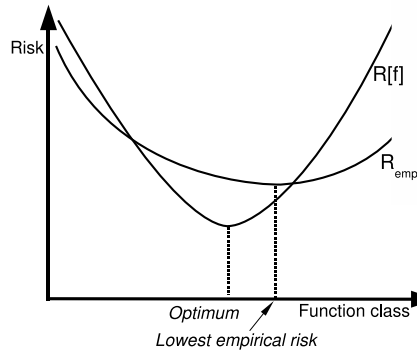}}
\caption{Empirical and expected risk as a function of the complexity.}
\label{figSRM}
\end{figure}

Usually it is possible to bring $R_{\mathrm{exp}}[f]$ to zero by choosing the classification function complex enough, where a measure for the complexity of a function 
will be introduced in the next section.
However choosing a function such that it fits the training examples perfectly is usually a typical case of overtraining (see \cite{basicsCite}). 
A plot of the complexity of the function against the expected
and empirical risk would therefore look somehow like in figure \ref{figSRM}. 
Thus the goal has to be to find a measure for the complexity of the function, in order to control it.
Then to find the optimal working point, which is usually not at zero empirical risk .
\subsection{Capacity of a function}
\begin{figure}
\centerline{\includegraphics[width=8cm]{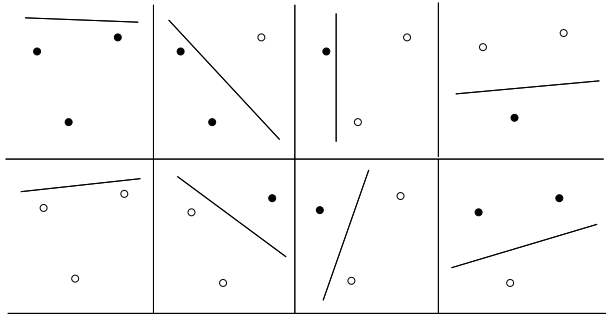}}
\caption{A line can separate three points independent of their labeling.}
\label{figVC}
\end{figure}

The complexity of a function can be measured by its capacity to separate points in two classes, independent of their labeling.
This concept has been introduced by Vapnik and Chervonenkis (VC) \cite{vapCheRef} and is thus known as the VC-dimension of a function.
Figure \ref{figVC} shows an example where the capacity of a line in a plane is illustrated. Since it is possible to separate
three points independently of their labeling, the VC-dimension is three.
Once the VC-dimension is known, another result from VC-theory can be used, namely the relation of the expected risk to the 
empirical risk and the capacity of the function:
\begin{equation}
R[f]\leq R_{emp}[f]+\Phi(h,m,\delta)
\end{equation}
This relation is valid with probability $1-\delta$, $m$ is the number of training examples and $h$ the VC-dimension.
$\Phi$ is a so called confidence term:
\begin{equation}
\Phi(h,m,\delta)=\sqrt{ \frac{1}{m}\left(h(\ln\frac{2m}{h}+1)+\ln\frac{4}{\delta}\right)}
\end{equation}

Obviously the limit on the estimation becomes worse if a higher probability is chosen, whereas
many training examples improve the limit.

\begin{figure}
\centerline{\includegraphics[width=7cm, trim=30 250 170 200, clip]{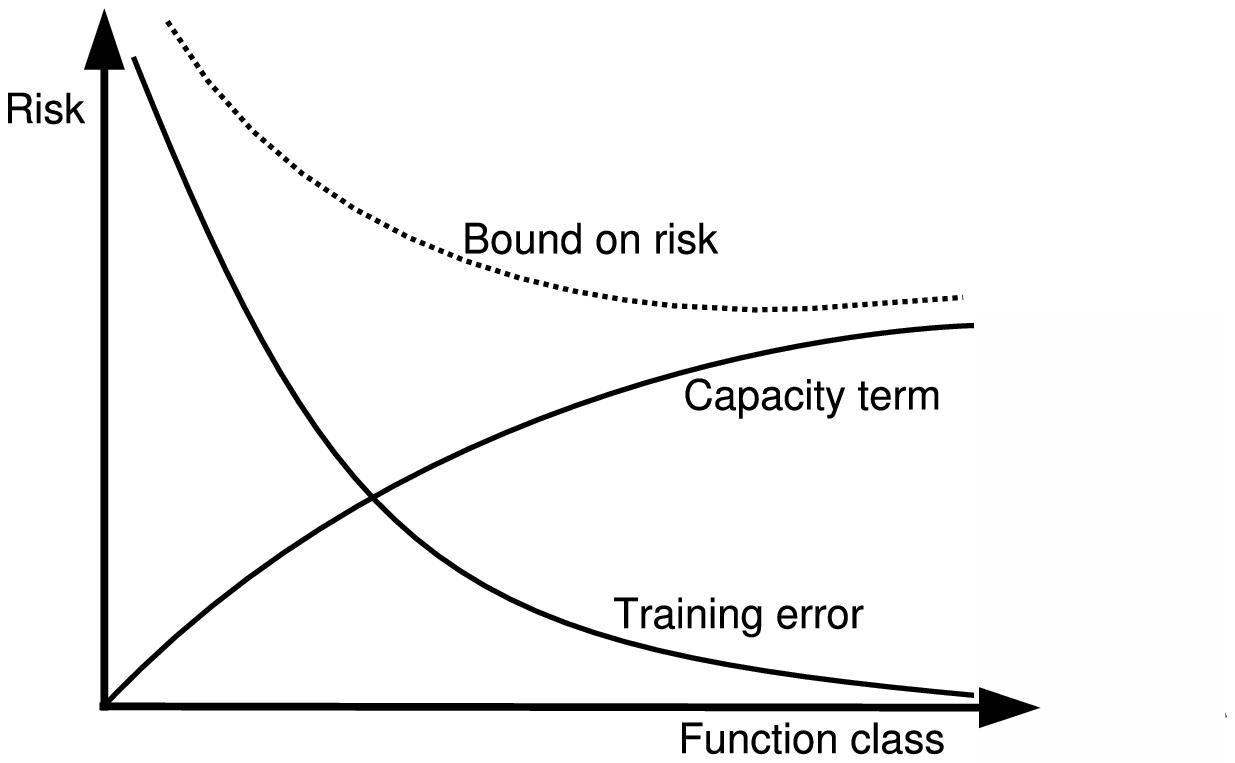}}
\caption{Bound on the expected risk derived from VC theory. Structural risk minimization tries to find the optimal point indicated.}
\label{figSRM2}
\end{figure}

A more complete treatment of the theory can be found in \cite{vapCheRef}. The dependence on the capacity 
term is shown in Fig. \ref{figSRM2}, where the dependence of the bound on the test error and of the training error 
is plotted exemplarily as a function of the VC dimension.
For ever more complex functions with higher VC-dimension, the training error will decrease, but 
the bound on the expected risk gets worse, because the capacity term grows.
Structural risk minimization thus deals with finding the minimum of the expected risk, 
by varying the complexity of the function. The evaluation can then be done as described in \cite{basicsCite}.

\subsection{Soft Margin SVM}

The Support Vector Machine has by construction a mean to control the capacity of the constructed classifier. 
That is the width of the margin. For details again see \cite{svmCite}.
The result that a wide margin corresponds to a function with low capacity, but good generalizing abilities.
In the linear case, this becomes clear from the arguments made in the beginning to motivate the maximum margin classifier.
But also after application of the kernel trick this stays true. The kernel trick implies a mapping from 
$H$ to $H'$ and it can be shown that a  wide margin in the $H'$ space corresponds to a smooth function in the input space $H$. 
However, the width of the margin is fixed by the support vectors, 
so in order to allow for a wider margin a higher empirical risk has to be accepted because the support vectors have to be allowed to lie then within the margin.
This idea leads to the so called Soft Margin SVM.
Here the target function in the Lagrangian (\ref{eqTau}) is modified to 
\begin{equation}
\label{targetFkt1}
\tau(\vec{w},\vec{\xi})=\frac{1}{2}|\vec{w}|^2+C\sum_{i=1}^m\xi_i
\end{equation}
The new variables $\xi_i \geq 0$ are chosen such that $y_i(<\vec{w},\vec{x}>)+b \geq 1-\xi_i $ for all $i =1,\ldots,m$ and are called slack variables.

\begin{figure}
\centerline{\includegraphics[width=8cm]{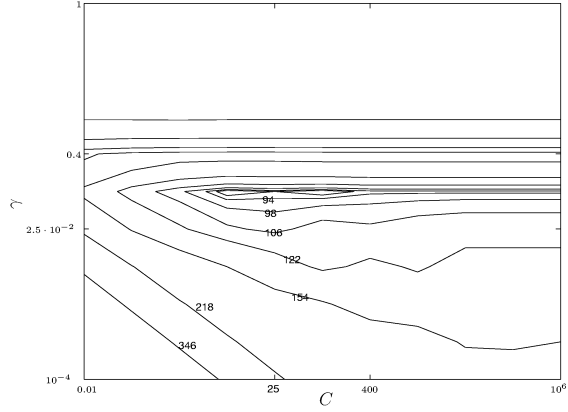}}
\caption{Exemplaric parameter space in logarithmic $C-\gamma$ coordinates for the classification of a dataset of characters (USPS), where each feature was
scaled between -1 and 1. The lines represent equal numbers of misclassified examples for this point in parameter space. This is in line with the
theoretical expectations \cite{asymBehaviour}.}
\label{figParSpace}
\end{figure}

In addition to the parameters that have to be optimized in the Lagrangians, and are determined by the SVM algorithm, 
there is now an additional parameter $C$ which controls the tradeoff between training error and capacity, which corresponds to the width of the margin.
Combined with the by far most popular kernel the gauss kernel, it leaves two parameters to be optimize by the user.
A typical picture of the $C$-$\gamma$ parameter space is shown in Fig. \ref{figParSpace}. 
The lines represent equal expected risk, estimated from an independent test set. This is just an example from a toy problem, 
but the simple layout is a general feature and has also a theoretical underpinning that leads to predictions in line with the shown
exemplaric case \cite{asymBehaviour}, \cite{chapelle}.
As already mentioned earlier, scanning this parameter space is in general no problem, and the search is stable.
Choosing optimal parameters is also known as model selection.

\section{Support Vector Regression}
Another slightly more general application of the above concepts is support vector regression (SVR).
For regression the goal is to construct a function with outputs $y\in\mathbb{R}$, that is a real valued function instead of $y\in\{\pm 1\}$.
Again, the linear case serves as a starting point, with the ansatz $f(\vec{x})=<\vec{w},\vec{x}>+b$ and the cost
incurred by the prediction $f(\vec{x})$ for the fv $\vec{x}$ is $c(\vec{x},y,f(\vec{x})):=y-f(\vec{x})_\epsilon:=max\{0|y-f(\vec{x})|-\epsilon\}$.
$||_\epsilon$ is the $\epsilon$-insensitive cost function introduced by Vapnik \cite{natureOfStatLearningTheory}.
By introducing a cost function that is insensitive to outliers with a distance smaller than $\epsilon$, 
the obtained solution has a sparse SV-expansion which is analog to the classification case.
This is because only examples on the margin become support vectors. 
With this, and after introducing slack variables $\xi_i$ and $\xi_i^*$ for the two cases
$f(\vec{x}_i)-y_i>\epsilon$ and $f(\vec{x}_i)-y_i<\epsilon$ the target function 
\begin{equation}
\tau(\vec{w},\vec{\xi},\vec{\xi^*})= \frac{1}{2}|\vec{w}|^2+C\sum_{i=1}^m(\xi_i+\xi_i^*)
\end{equation}
is built.
Again, analogous to Eq. (\ref{targetFkt1}), this function has to be minimized, leading to a support vector expansion
for the function $f(\vec{x})$ that reads, with the kernel trick already incorporated, 
\begin{equation}
f(\vec{x})=\sum_{i=1}^m(\alpha_i^*-\alpha_i)k(\vec{x}_i,\vec{x})+b
\end{equation}
where $k(\vec{x}_i,\vec{x})$ is the kernel function and the $\vec{x}_i$ the support vectors.
For details again see \cite{svmCite}. 
Although currently not used very much, there have been very encouraging results using support vector regression (e.g. \cite{bostonHousing}) 
that again show the power of kernel methods.

\section{Tools}
\label{toolSect}
To learn about algorithms it is always necessary to build a system and try them on different datasets to get a feeling for what 
works and what doesn't work. 
In this section some tools are presented that facilitate this task. 
First the weka workbench \cite{weka}, that was already mentioned in \cite{basicsCite}. This is a veritable workbench, 
which has the advantage that it is possible to go through the whole pattern recognition process from feature selection to classification.
The interplay between SVMs and the selected and preprocessed features can be tried. 
To incorporate SVMs in the code, the ``libsvm''  library developed by Chih-Chung Chang and Chih-Jen Lin \cite{libsvm} can be used.
This is a popular C-library that also has a graphical user interface and a good documentation. 
Also available is the very similar, but based on C++ templates, implementation of Olaf Ronneberger, which is based on the code 
of the libsvm, but uses more modern interfaces \cite{libsvmTL}.
Also available now is an implementation in the Toolkit for Multivariate Analysis TMVA \cite{tmva}, which is incorporated 
into  ROOT.
Here the advantage is that ROOT is widely used in HEP, so there is no need to incorporate a new tool. TMVA is easy to use,
has also a graphical user interface. It can be used from the users code via a C++ or a scripting interface. Furthermore it is adapted to the use 
in particle physics analysises.
So, there are enough possibilities to try out these new concepts. 

\section{Applications in HEP}
To show the usefulness of the above concepts in data analysis for high energy physics, two results from 
the application of SVMs to HEP analysis are presented.
The first is a top quark analysis at the Tevatron. The authors of \cite{topAtTev} tried to improve the signal to background ratio
of the dilepton channel, with a branching ratio of about 5\%, to identify $t\bar{t}$ events. Since a signature of an interesting event 
are two high $p_T$ leptons, large missing $E_T$ and two $b$-quark jets, variables that contain these informations, namely lepton and first jet
$E_T$, the sum of the jet $E_T$ and the missing $E_T$ are used as input. 
Combined with a standard libsvm SVM implementation trained on precut simulated  Monte Carlo data using a gaussian kernel, this approach outperformed optimized cuts.
The discriminating power of the variables and the signal efficiency vs. background efficiency can be seen in \cite{topAtTev}.

\begin{figure}
\centerline{\includegraphics[width=8cm,trim=0 100 0 0,clip]{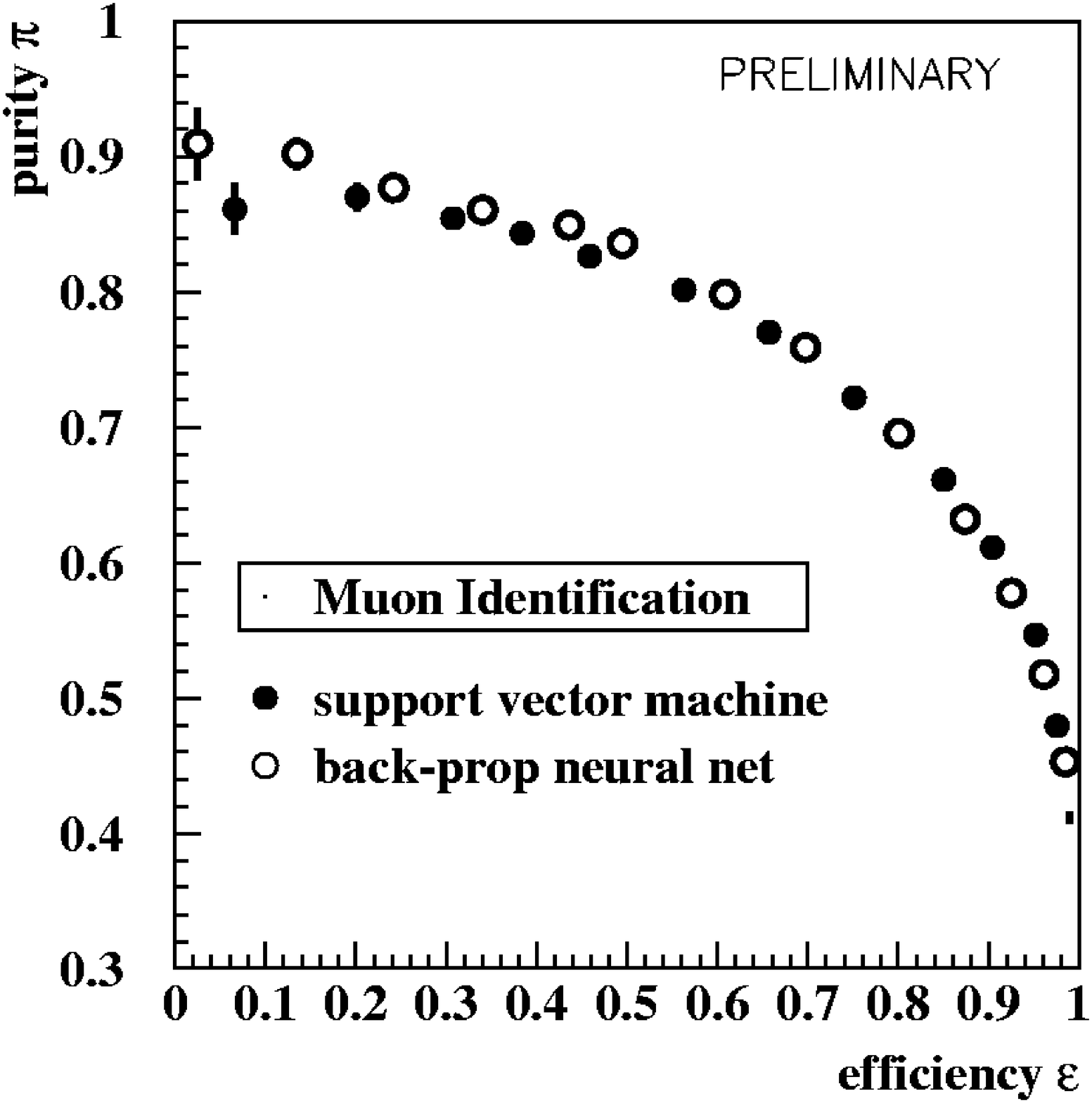}\includegraphics[width=6.2cm]{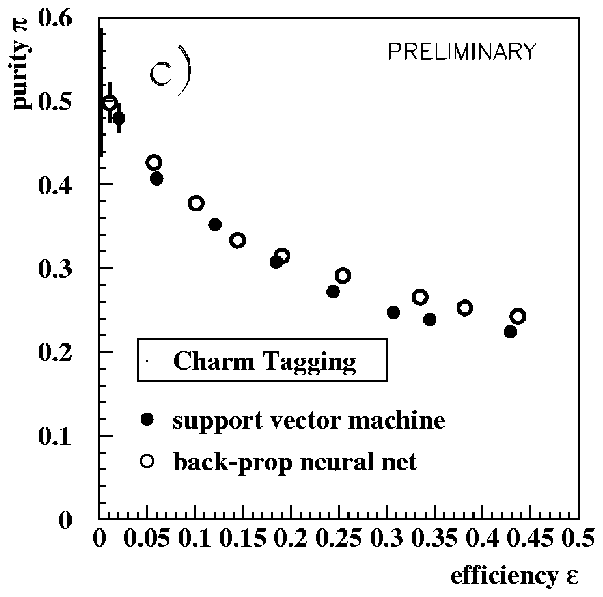}} 
\caption{Results for the comparison between neural networks and SVMs from a flavour separation problem from the OPAL experiment at LEP 
\cite{opalSVM}}
\label{figOpal}
\end{figure}

The second example for the successful application of SVMs in HEP comes from the OPAL experiment at LEP where the use of SVMs for 
flavour tagging and muon identification was investigated \cite{opalSVM}.
Here rather complex input variables were chosen, based on their discriminating power and a special emphasis was put on the comparison 
with a neural network, that was trained with the same data.
 Fig. \ref{figOpal} shows the performances of the SVM compared with the neural network.
It is obvious that the performance of the SVM is equal to that of the NN even though the SVM approach is fairly new and some input 
variables where chosen based on their performance with the NN. For further comparisons between NNs and SVMs, see \cite{svmVsNN}.


\section{Conclusion}
After the presentation of the basics of support vector machines,tools and results, this conclusion includes some advice when and why to use SVMs.
Firstly, it is probably not possible to replace an existing neural network algorithm that works fine with a SVM to get directly better performance.
NN are still performing at least as good as out of the box SVM solutions in HEP problems, plus they are usually faster if the data is noisy and the number
of support vectors is therefore high. 
This is a specific problem in HEP and it should be addressed for example in the preprocessing step.
Another point of importance in HEP, which has to be clarified, is the robustness against training data that does not exactly reproduce 
real data. 
On the other hand, SVMs are a new tool for HEP that is easy to use, even for a non expert. 
So it is worth trying them on new problems and compare the results with other out of the box multivariate analysis methods. 
The advantage of the SVM is that they are theoretically understood, which also opens the way for new, interesting developments,
and if more people are using them in physics, new solutions will maybe lead to classifiers that are better than today's algorithms.
It has to be kept in mind, that  neural networks, now common in data analysis, took quite some time to become accepted and tailored for HEP problems.
Furthermore the kernel trick is a general concept. Having experience with kernels suitable for SVMs in HEP opens the possibility to use other kernel methods like 
the ones mentioned in section \ref{kernelTrickSect}.
For those who want to know more about tools, documentation and publications on this topic a good starting point is the website www.kernel-machines.org.

\end{document}